**Towards Accurate and Fast Clinical Body Composition: A Resource-Efficient Hierarchical Segmentation Framework for Multi-Source CT**

Xiaodi Shen[1†], Qingzhu Zheng[1†], Yaoyang Qiu[2†], Cien Fan[1], Ruonan Zhang[1], Yangdi Wang[1], Luyao Wu[1], Weikai Zheng[1], Longfei Zhao[2], Bing Li[2], Rulin Xu[2] , Qiqi Xu[2*], Ren Mao[1*], Shiting Feng[1*], and Xuehua Li[1*]

[1] Department of Radiology, The First Affiliated Hospital, Sun Yat-sen University, Guangzhou 510080, China

[2] Research & Development Center, Canon Medical Systems (China) Co. Ltd. Beijing 100015, China

[†] These authors contributed equally to this work.

[*] Corresponding authors: qiqi.xu@cn.medical.canon (Qiqi Xu); maor5@mail.sysu.edu.cn (Ren Mao); fengsht@mail.sysu.edu.cn (Shiting Feng); lxueh@mail.sysu.edu.cn (Xuehua Li)

## Abstract

**Background:** Automated volumetric segmentation of skeletal muscles and adipose tissue from Computed Tomography (CT) is critical for large-scale body composition analysis. However, deploying standard deep learning architectures across heterogeneous clinical environments remains challenging due to performance degradation on multi-source data and high computational/memory demands during CPU-based inference.

**Methods:** Leveraging a multi-institutional dataset, we propose a coarse-to-fine hierarchical framework for segmenting ten tissue structures. To enhance efficiency, a Dynamic Spacing and Anisotropic Patching strategy is employed across both stages to capture global layouts and reduce patch numbers by adapting to voxel anisotropy. In the fine stage, a Group Inference mechanism enables incremental local processing during sliding-window inference, releasing prediction sub-volumes on-the-fly to minimize peak memory. Additionally, a Topology-Aware Asymmetric Resampling algorithm is introduced to accelerate post-processing by selectively interpolating boundary and interior voxels.

**Results:** The framework was trained on 1,558 CT volumes from seven public and two private datasets, and evaluated on an independent test cohort ( $N = 105$ ). Per-structure Dice Similarity Coefficients (DSC)—evaluated on target-specific sub-cohorts ($N = 20{\sim}50$) based on anatomical coverage—ranged from $0.924 \pm 0.025$ (intercostal muscles) to $0.982 \pm 0.011$ (subcutaneous fat) across the ten targets. Bland–Altman analysis indicated that the 95% Limits of Agreement fell within the predefined $\pm 10\%$ relative error clinical acceptance limit for eight major structures; the remaining two thin-line structures (biceps femoris and intercostal muscles) exhibited marginal variances due to their inherent geometric sensitivity to boundary relative error calculations. On a 12-core Intel Xeon workstation utilizing the OpenVINO runtime, the pipeline demonstrated a mean inference time of 44.5 seconds per volume with a mean peak memory of 4.73 GB.

**Conclusion:** By integrating dynamic patch adaptation, incremental group inference, and topology-aware resampling into a hierarchical pipeline, our framework achieves an optimal balance between segmentation accuracy and computational efficiency. It provides a robust, GPU-free paradigm that facilitates routine, large-scale body composition analysis on standard CPU workstations across heterogeneous clinical workflows.

**Keywords:** Body Composition Analysis, Coarse-to-Fine Framework, Group Inference, Dynamic Patching, Topology-Aware Resampling.

## 1. Introduction

### 1.1. Clinical Background and Technical Challenges

Quantitative body composition analysis from Computed Tomography (CT) images—specifically measuring skeletal muscle and adipose tissue compartments—has become a cornerstone for metabolic health screening [1-3], cancer cachexia monitoring [4-6], and sarcopenia assessment [7-9]. Visceral adipose tissue (VAT) accumulation, in particular, serves as a critical pathogenic driver of metabolic syndrome, type 2 diabetes, and cardiovascular diseases [10-12]. However, translating deep learning-based segmentation frameworks from well-curated academic benchmarks into resource-constrained clinical pipelines remains non-trivial. Real-world deployment is bottlenecked by severe scan heterogeneity (e.g., highly anisotropic slice thicknesses), strict CPU-only memory limits on clinical workstations, and the morphological complexity of boundary-sensitive, small-volume structures like intercostal muscles.

Despite the rapid evolution of large-scale foundational networks, existing methodologies fail to fully reconcile the trilemma of inference speed, memory efficiency, and boundary fidelity in production environments:

- **Computational Redundancy and Spatial Inflexibility:** Large-scale models typically rely on rigid, isotropic patch-based inference. When encountering real-world CT data with highly anisotropic voxel spacing, this static partitioning introduces severe geometric domain mismatch. Furthermore, frameworks like TotalSegmentator [14] distribute distinct anatomical targets (e.g., individual muscles versus adipose tissue) across separate, independent sub-models, which drastically multiplies inference latency and computational redundancy.
- **Prohibitive Memory Bottlenecks on CPU Hardware:** For comprehensive 10-class body composition panels, standard sliding-window protocols mandate the accumulation of high-dimensional prediction probability tensors across all overlapping patches before executing global post-processing. This retention mechanism generates a peak memory footprint that scales linearly with the class count and volume size, routinely exceeding the strict RAM budgets of standard clinical CPU workstations and triggering out-of-memory crashes.
- **Efficiency-Fidelity Tradeoff in Resolution Recovery:** Current models fail to preserve the fine-grained boundary fidelity required for accurate volumetric tracking. To achieve high-precision anatomical boundaries during resolution recovery, the theoretically optimal approach is to apply linear interpolation directly onto continuous probability maps. However, to maximize runtime performance and minimize computational overhead, current production frameworks, like notably TotalSegmentator and VibeSegmentator [18], compromise this fidelity by converting outputs into discrete binary masks prior to executing low-overhead nearest-neighbor interpolation. This operational shortcut inherently discards crucial sub-voxel boundary details and induces severe partial volume effects, blurring thin-line anatomical interfaces (e.g., intercostal muscles) and translating into non-negligible volumetric calculation errors.

Consequently, practical clinical deployment invites a shift toward application-driven pipeline optimization. This necessitates a standardized data workflow that simultaneously addresses inference efficiency, memory-constrained environments, and post-processing latency without imposing additional online computational overhead.

### 1.2. Core Contributions of This Work

To bridge the gap between academic body composition research and production-grade clinical utility, this paper presents an optimized, deployment-ready coarse-to-fine hierarchical segmentation system targeting ten distinct body composition structures (encompassing key adipose tissue compartments, major muscle categories, and individual muscle groups) from heterogeneous CT data. The primary contributions of this study are summarized into three independent and complementary dimensions:

1. **Inference-Accelerated Hierarchical Pipeline with Budget-Constrained Anisotropic Patching:** We develop a robust hierarchical workflow that pairs a rapid multi-structure localization network with a high-resolution sub-region refinement network. To optimize inference latency, the framework incorporates a **Dynamic Spacing and Anisotropic Patching Strategy**. Guided by the scale-tolerance bounds derived from training-time augmentations, the algorithm dynamically adjusts the aspect ratio and dimensional spacing of the sliding-window patch (e.g., restructuring an isotropic $160 \times 160 \times 160$ volume into an anisotropic $128 \times 160 \times 192$ configuration). This strategy maximizes single-patch coverage along the dominant anatomical axis, substantially reducing the required number of sliding-window iterations and compressing inference latency while maintaining strict accuracy equivalence.
2. **Deployment-Driven Optimization for Peak Memory and Post-Processing Efficiency:** To address the resource limitations of clinical CPU workstations, we introduce a dual-optimization protocol that reduces runtime resource consumption without sacrificing segmentation accuracy:
    - *(i) Group Inference Protocol:* By executing incremental local processing during the inference sweep, the pipeline dynamically identifies and finalizes completed sub-volumes on-the-fly, immediately releasing the associated high-dimensional softmax tensors and significantly mitigating peak memory accumulation.
    - *(ii) Topology-Aware Asymmetric Resampling:* In the resolution recovery phase, voxels are classified into anatomical core regions (processed via low-latency zero-order mapping) and high-risk boundary zones (processed via first-order linear interpolation), achieving a near-linear speedup in post-processing while preserving boundary smoothness.
3. **A Rigorous Multi-Center Clinical Validation and Volumetric Agreement Analysis:** The clinical utility and geometric robustness of the framework are validated through a comprehensive evaluation paradigm. Trained on 1,558 volumes and evaluated on an independent test cohort ($N = 105$), our framework demonstrates strict clinical equivalence to expert manual annotations. Through Bland–Altman analysis against predefined Clinical Acceptance Limits ($\pm 10\%$ relative error), we establish the reliability of our automated volumetric measurements and provide an in-depth clinical analysis of geometric boundary sensitivity in thin-line anatomical structures (e.g., intercostal muscles).

## 2. Methodology

### 2.1. Data Infrastructure and Quality Control

#### 2.1.1. Dataset Construction and Multi-Source Heterogeneity

To enhance the system’s geometric generalizability and robustness against multi-center domain variations, the baseline imaging repository was aggregated from a combination of private clinical registries and dominant public benchmarks. This heterogeneous data pool incorporates diverse CT volumes harvested from multi-site institutional archives alongside established academic cohorts, including the Medical Segmentation Decathlon (MSD: Task03 Liver, Task07 Pancreas, Task08 Hepatic Vessel, Task09 Spleen) [13], TotalSegmentator [14], C4KC-KiTS [15], TCGA-LIHC [16], and CTpred-Sunitinib-panNET [17].

Following a general programmatic quality screening to ensure basic contrast viability and artifact clearance, a total of 1,663 high-quality 3D volumes encompassing a wide spectrum of healthy controls, oncological patients, and mixed clinical populations were stabilized. To support a rigorous industrial train-test split, this aggregated registry was partitioned at the patient level into an Internal Training Cohort (1,558 cases) dedicated to offline model optimization, and an Independent Static Test Cohort (105 cases) reserved exclusively for terminal non-contact production testing.

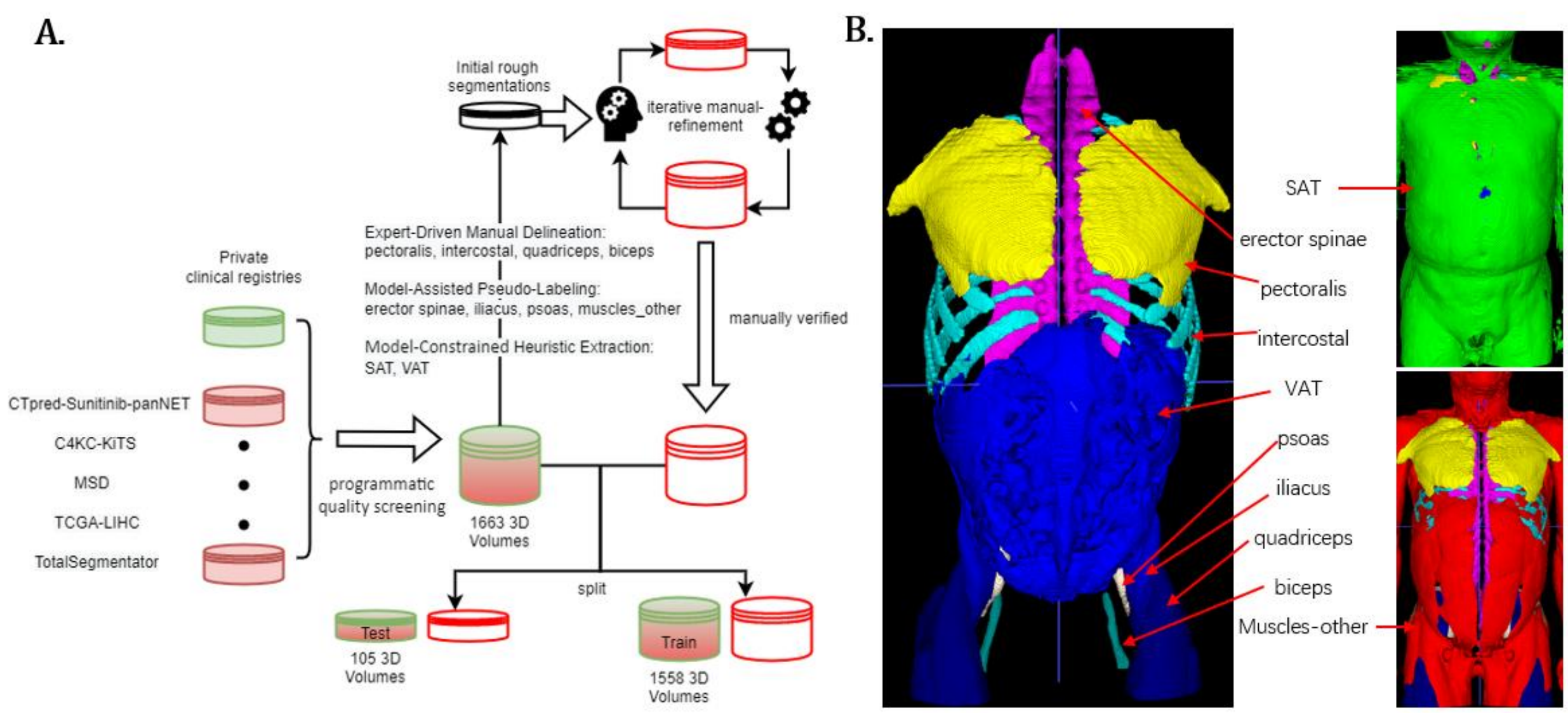

Figure 1. Schematic diagram of the dataset workflow and segmentation targets. (A) Data pipeline and expert-in-the-loop quality control: Illustration of multi-source heterogeneous data aggregation, cohort partitioning (N=1,558 for training and N=105 for testing), and the three-stream iterative annotation protocol (manual, model-assisted, and heuristic extraction with closed-loop refinement). (B) Exemplary target masks: Demonstration of the definitive semantic labels for the ten musculoskeletal and adipose tissue components on representative CT views.

As shown in Table 1 and Figure 2, the training corpus exhibits deliberate, significant heterogeneity across three orthogonal axes: (i) **Scanner vendor**: multiple CT manufacturers and diverse hardware generations; (ii) **Acquisition protocol**: high variability in slice thickness, in-plane pixel spacing, tube current, and tube voltage; (iii) **Demographic distribution**: a widespread coverage of age groups and sex ratios, encompassing broad clinical physiological spectrums

Table 1: Baseline characteristics of the dataset

| Characteristic | Distribution |
|---|---|
| Case Number (Train / Test) | 1663 (1558 / 105) |
| **Scanner vendor** | |
| Manufacturer | Unknown: 1239; Canon: 207; Siemens: 155;<br>GE: 28; Philips: 34 |

| **Acquisition protocol** | |
|---|---|
| Tube Current (mA) | 284.33 ± 125.74 |
| KVP (top 3) | 120kV: 374; 100kV: 34; 140kV: 10 |
| Slice Thickness (mm) | 2.91 ± 1.72 [0.5, 7.0] |
| Pixel Spacing (mm) | 1.07 ± 0.36 [0.3, 1.5] |
| **Demographic distribution** | |
| Age (years) | 56.48 ± 15.21 [6.0, 91.0] Available 313/1663 |
| Sex (M / F / Unknown) | 177 / 140 / 1346 |

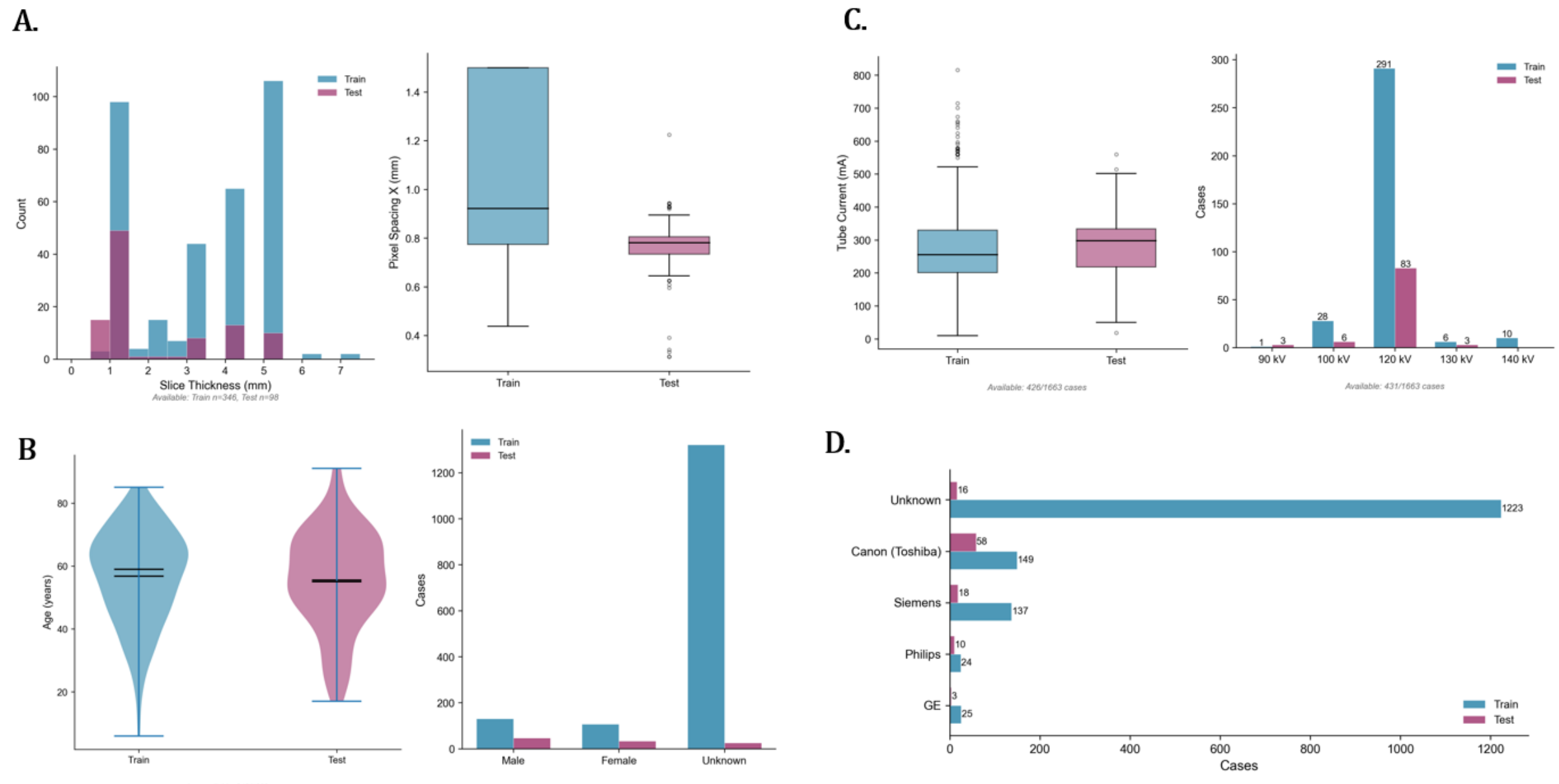

Figure 2. Statistical profiles illustrating multi-center dataset heterogeneity. (A) Distributions of longitudinal slice thickness and transaxial in-plane pixel spacing. (B) Variations in tube current (mA) and peak tube voltage (kVp) configurations. (C) Demographic breakdown across age cohorts and biological sex distributions. (D) Proportional representation of different hardware scanner manufacturers across the aggregated repository.

### 2.1.2. Annotation Protocol and Ground Truth Generation

To ensure the anatomical accuracy and clinical validity of the target masks, a rigorous, expert-in-the-loop iterative annotation protocol was established to compile the ground-truth annotations. Rather than applying a uniform segmentation baseline, initial rough contours were generated via a hybrid, category-specific initialization strategy meticulously tailored to the tissue contrast and spatial characteristics of each structure:

- **Expert-Driven Manual Delineation (Scratch Initialization):** For anatomically intricate muscle groups that lack highly accurate representation in public benchmarks or present high surface-to-volume ratios—specifically the pectoralis, intercostal, quadriceps femoris, and biceps femoris muscles—imaging specialists manually delineated a minority of pilot cases from scratch to establish a bias-free geometric baseline.
- **Model-Assisted Disaggregation and Residual Extraction:** For standard structures and broader

muscular compartments, initial pseudo-labels were derived using automated networks. The erector spinae was directly imported via the pre-trained TotalSegmentator engine. Conversely, the psoas major and iliacus (natively grouped as a single class in TotalSegmentator) underwent an expert-guided topological decoupling to separate them into distinct anatomical entities. Since TotalSegmentator predictions often suffer from over-smoothing and overly conservative (under-segmented) boundaries, rigorous expert-guided refinement was performed to restore anatomically precise outer contours. Furthermore, to capture the comprehensive muscle bulk, a dedicated network was trained on a commercially viable subset of the SAROS [19] dataset; a residual class, designated as “muscles_other”, was subsequently extracted by pixel-wise subtracting the individually tracked core muscles from this bulk mask to serve as a formal training target.

- **Model-Constrained Heuristic Extraction:** For fat compartments, initial localization was governed by the SAROS-trained model to define the broad subcutaneous and intra-abdominal spatial envelopes, with the gastrointestinal tract explicitly masked out based on TotalSegmentator references to avoid partial volume contamination. Within these model-restricted regions, the final initial boundaries for subcutaneous adipose tissue (SAT) and visceral adipose tissue (VAT) were derived by applying a standardized Hounsfield Unit (HU) threshold window of [-190, -30] coupled with localized morphological operations.

During downstream quantitative clinical evaluations, the “muscles_other” class was actively reserved for fusion with the individual muscle masks to assess the comprehensive “muscles_all” macro-compartment.

To eliminate systematic machine biases and subtle boundary discrepancies across all core target classes, the initialized masks from the three streams were subjected to a rigorous, closed-loop manual-refinement pipeline. Subsets exhibiting anatomical deviations were manually corrected by specialists and fed back to retrain local networks iteratively until all target contours achieved complete topological consistency. Finally, to eliminate intra-observer variability and establish an absolute gold standard, 100% of the terminal segmentations were audited and formally verified by a senior annotation engineer possessing 8 years of specialized experience in clinical radiology.

## 2.2. Overall Hierarchical Pipeline

The proposed body composition segmentation framework adopts a structured coarse-to-fine hierarchical architecture designed to balance high multi-structure annotation accuracy with stringent inference efficiency constraints. To streamline the data workflow and eliminate complex class-mapping translations between different stages, both levels of the hierarchy are configured to output an identical set of semantic labels, maintaining structural pipeline uniformity. As illustrated in Figure 3, the overall workflow progresses through two primary operational phases:

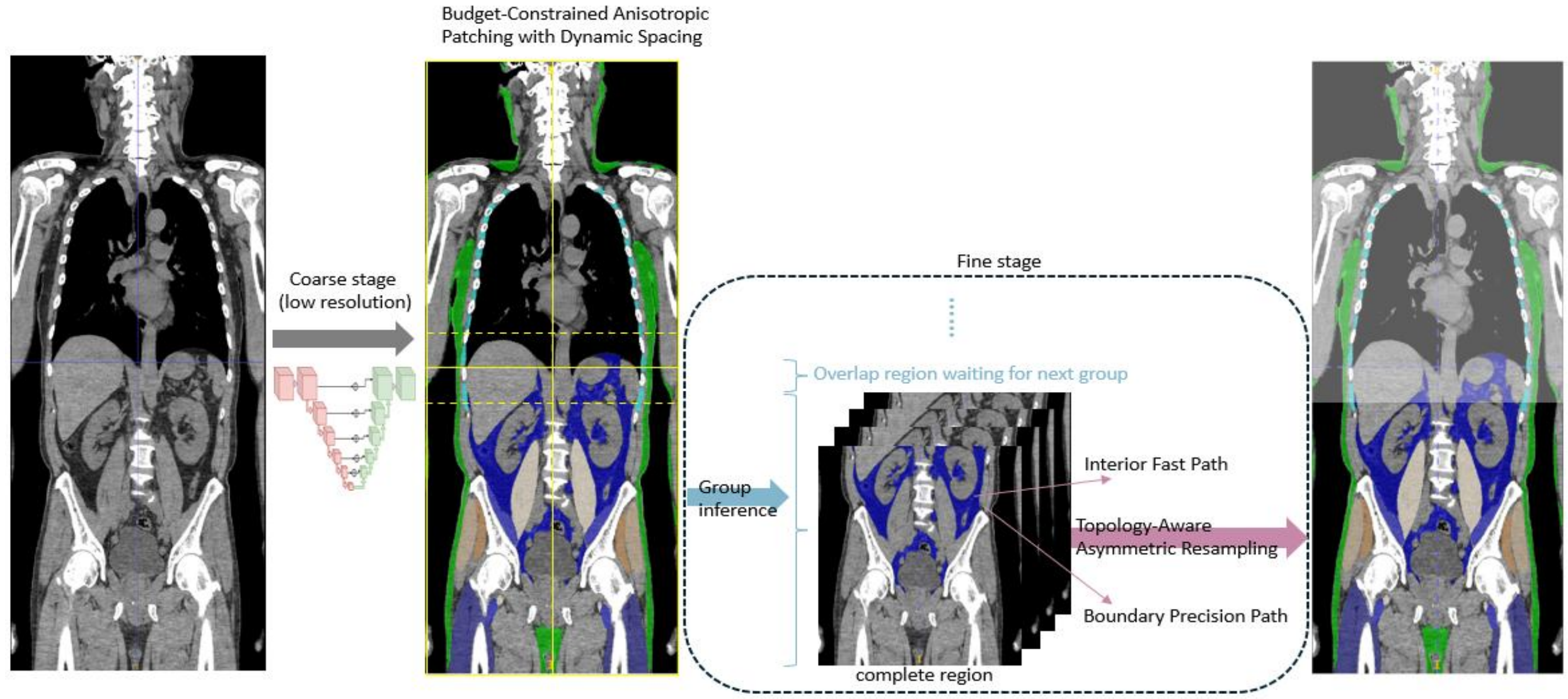

Figure 3. Schematic overview of the proposed coarse-to-fine hierarchical segmentation framework. The corporate pipeline consists of two primary sequential phases: (1) The Coarse Stage, which operates at a lower resolution to dynamically eliminate non-target artifacts (e.g., scan beds, air) and bound a 3D Region of Interest (ROI); and (2) The Fine Stage, which performs high-resolution dense volumetric inference within the localized ROI to segment all ten tissue structures, accelerated by memory-efficient Group Inference and a Topology-Aware Asymmetric Resampling protocol. Both stages maintain identical semantic label outputs to ensure structural pipeline uniformity.

**1. Coarse Stage (Global Volume Reduction):** The input CT volume is processed at a lower resolution to efficiently identify and eliminate computationally redundant non-target regions, such as scanning beds, surrounding air, and irrelevant anatomical structures. This stage dynamically delineates a clean, optimized 3D Region of Interest (ROI) encompassing all valid target tissues.

**2. Fine Stage (Targeted ROI Refinement):** Within the constrained spatial envelope defined by the coarse ROI, a high-capacity network performs dense, high-resolution volumetric inference to end-to-end segment all individual muscle groups, fat compartments, and auxiliary classes. The execution of the fine stage is fortified by two integrated mechanisms: memory-efficient Group Inference and a Topology-Aware Asymmetric Resampling protocol.

## 2.3. Coarse Stage: Efficient Multi-Structure Localization

The coarse model leverages a standard 3D U-Net topology with a compact channel specification of $[16,32,64,128,160,160]$ across six encoder-decoder stages, operating on patches of $224 \times 128 \times 128$ voxels at a base training spacing of $(3.8,3.04,3.04)$ mm. This lightweight configuration is optimized for high-throughput global localization and background removal. To minimize computational overhead and eliminate the need for multi-patch sliding-window inference at this stage, we introduce a Budget-Constrained Anisotropic Patching Strategy.

### 2.3.1. Budget-Constrained Anisotropic Patching with Dynamic Spacing

Standard convolutional networks typically exhibit performance degradation when subjected to substantial resolution variations during inference. However, by introducing scaling augmentations during the training phase, the network develops intrinsic resilience to varied spatial spacings within a bounded tolerance range.
To exploit this property for inference acceleration, the system adaptively calculates the maximum permissible voxel spacing and its corresponding anisotropic patch aspect ratio based on the boundary limitations derived from training-stage data augmentations. The inference spacing $s$ is scaled relative to the training spacing $s_{\text{train}}$ by a scaling factor $\alpha$:

$$s = \alpha \cdot s_{\text{train}}, \quad \alpha \in [\alpha_{\text{min}}, \alpha_{\text{max}}]$$

For the coarse model, $\alpha_{\text{min}} = 1.0$ and $\alpha_{\text{max}} = 1.4$, with $s_{\text{train}} = (3.8, 3.04, 3.04)$ mm—the upper bound being directly inherited from the scale augmentation range applied during training.
Crucially, to preserve the physical receptive field under varying slice thicknesses without exceeding the hardware memory capacity, the discrete patch dimensions $P = (p_x, p_y, p_z)$ are dynamically adapted. Given a maximum voxel computational budget $V_{\text{max}}$ (equivalent to the standard training volume envelope, e.g., $160 \times 160 \times 160$), the patch size is reformulated as an anisotropic tensor under the rigid bound constraint:

$$\prod_{d \in \{x,y,z\}} p_d \leq V_{\text{max}}$$

where the ratio $p_x : p_y : p_z$ is continuously adjusted proportional to the physical anisotropy of the input CT data. By dynamically expanding the spacing and adaptively restructuring the patch aspect ratio (e.g., transforming a baseline isotropic patch into an asymmetric $128 \times 160 \times 192$ volume), the entire region of interest can be processed within a single inference patch for the majority of cases. This eliminates redundant patch-stitching artifacts and reduces unnecessary sliding-window iterations.

## 2.4. Fine Stage: Targeted Multi-Structure Refinement

The fine segmentation model shares the foundational 3D U-Net backbone of the coarse stage but incorporates an expanded model capacity—with channel dimensions $[32, 64, 128, 256, 320, 320]$—and operates on higher-resolution ROI inputs with a base training spacing of $(1.5, 1.2, 1.2)$ mm and patch size of $160 \times 160 \times 160$ voxels. the Budget-Constrained Anisotropic Patching Strategy introduced in Section 2.3.1 is symmetrically deployed in this stage, to ensure high anatomical fidelity, the dynamic spacing range is restricted to $\alpha \in [1.0, 1.1]$.
To optimize the model for clinical resource limitations and boundary blurring without altering the core representation layer, the fine stage integrates two complementary strategies: Group Inference and Topology-Aware Asymmetric Resampling.

### 2.4.1. Group Inference: Memory-Efficient Incremental Post-Processing

A primary bottleneck in multi-class 3D segmentation arises from peak memory consumption during sliding-window inference. The total runtime memory footprint $M_{\text{total}}$ can be modeled as:

$$M_{\text{total}} = M_{\text{network}} + M_{\text{results}}$$

where $M_{\text{network}}$ denotes the memory occupied by the network's activations during a single-patch forward pass, and $M_{\text{results}}$ corresponds to the accumulation of softmax (or logit) prediction tensors for all patches covering the volume. Critically, $M_{\text{results}}$ scales linearly with the product of the total number of voxels and the number of target classes. In comprehensive body composition analysis involving multiple output classes over an extended scan range, $M_{\text{results}}$ often exceeds $M_{\text{network}}$, becoming the dominant contributor to out-of-memory errors on standard clinical workstations.

To mitigate this, we propose **Group Inference**—a memory-efficient incremental scheme that restructures the sliding-window processing sequence. Rather than deferring all post-processing until every patch has been processed, completed prediction sub-volumes are identified on-the-fly during the inference sweep. The algorithm operates through the following pipeline:

1. During sliding-window traversal, the system dynamically tracks spatial sub-regions that have received contributions from all overlapping patches.
2. Once a sub-region is confirmed as complete—meaning no remaining unprocessed patches overlap with its spatial coordinates—a local post-processing operation (argmax) is immediately executed on that specific sub-volume.
3. The high-dimensional softmax tensor for the completed sub-region is converted into a compact integer label map and immediately released from active memory.
4. Steps 2–3 repeat incrementally as the sliding window advances, maintaining only the softmax tensors for actively accumulating sub-regions.

The peak memory under Group Inference is strictly bounded by:

$$M_{\text{peak}}^{\text{group}} \approx M_{\text{network}} + \max_{k}\left(\sum_{i\in\mathcal{A}(k)} M_{\text{softmax}}^{(i)}\right)$$

where $\mathcal{A}(k)$ indexes the set of spatial sub-regions whose predictions remain incomplete at step $k$. Since completed regions are progressively finalized and freed, the active accumulation set $\mathcal{A}(k)$ remains small relative to the full volume. This strategy controls peak memory consumption to be comparable to single patch inference, enabling multi-class body composition analysis within standard clinical hardware constraints with zero mathematical loss of accuracy.

#### 2.4.2. Topology-Aware Asymmetric Resampling

The conversion of model predictions from inference resolution back to native CT resolution constitutes a mandatory post-processing step. In production environments targeting CPU-only execution, standard linear interpolation over the entire prediction volume presents a noticeable computational overhead.

We exploit a structural property of segmentation predictions to achieve computational acceleration: the interior voxels of a predicted structure—defined as those whose spatial neighbors all share the identical predicted class—can be mapped via efficient nearest-neighbor interpolation with zero discretization error. Only boundary voxels residing at tissue interfaces require the precision of linear interpolation to preserve smooth anatomical contours.

The **Topology-Aware Asymmetric Resampling** algorithm formalizes this observation into a dual-path reconstruction workflow:

1. **Boundary Classification:** The coarse prediction at inference resolution is scanned to classify each voxel as *interior* (all spatial neighbors within a defined kernel share the same predicted class) or *boundary* (at least one neighbor differs).
2. **Interior Fast Path:** Interior-classified voxels are upsampled to the native CT resolution via constant nearest-neighbor interpolation, incurring minimal computational cost.

3. **Boundary Precision Path:** Boundary-classified voxels are upsampled via standard linear interpolation to preserve sub-voxel contour fidelity.
4. **Masked Fusion:** The two independently upsampled label maps are merged using the upsampled boundary mask, producing a unified full-resolution segmentation.

Since most voxels in target structures reside within homogeneous interior regions, this dual-path strategy reduces the computational workload to the boundary pixels. The approach introduces zero accuracy degradation by construction: interior voxels are mathematically guaranteed to produce identical results under both interpolation modes.

### 2.5. Pre-processing and Post-processing Pipelines

#### 2.5.1. Standardized Pre-processing Workflow

**Coarse Input:** The full CT volume is first analyzed to determine its physical bounding box dimensions. The inference spacing and patch size are then adaptively computed by the Budget-Constrained Anisotropic Patching module (Section 2.3.1). The volume is resampled to the determined spacing, followed by Hounsfield Unit clipping and z-score intensity normalization conforming to standard configurations.

**Fine Input:** Based on the multi-structure localization mask produced by the coarse stage, the surrounding uninformative regions (e.g., external scanning air, table artifacts) are cropped out. A tight, optimized bounding box encompassing the remaining target body bulk is extracted. The cropped sub-volume is resampled and normalized to match the fine-stage resolution requirements.

#### 2.5.2. Deployment Post-processing

Following high-resolution inference in the fine stage, the Topology-Aware Asymmetric Resampling algorithm (Section 2.4.2) is applied to recover the native CT resolution. Standard connected-component analysis (CCA) is optionally applied to eliminate isolated false-positive voxels.

The fine network natively outputs ten discrete semantic target classes—including the individual target muscles, SAT, VAT, and the “muscles_other” residual compartment. During production and deployment, these masks are directly retained for clinical representation. Crucially, to accommodate macro-level volumetric metrics during downstream clinical evaluations, a rule-based post-hoc evaluation pipeline is executed: the individual core muscle masks and the network-predicted “muscles_other” mask are pixel-wise aggregated to dynamically synthesize the comprehensive “muscles_all” macro-compartment. The final multi-label mask is resampled to the original CT coordinate space for terminal presentation.

## 3. Experiments and Results

### 3.1. Implementation and Training Details

The coarse stage localization network utilizes a lightweight 3D U-Net topology with an initial feature channel depth of 16, optimized for high-throughput global localization. The fine stage refinement network expands the backbone parameter space with an initial feature depth of 32, operating on dynamically cropped high-resolution patches.

Both models were trained using the Stochastic Gradient Descent (SGD) optimizer with an initial learning rate governed by a polynomial decay schedule. The compound loss function combines Auto-Weighted Cross-Entropy (with boundary emphasis factor $\gamma$) and Batch Dice Loss. Training proceeded for 1,000 epochs with a batch size of 2.

All model training was implemented on top of the open-source nnU-Net [20] framework (built upon PyTorch), leveraging enterprise GPU computing infrastructure. The network was regularized using nnU-Net's native data augmentation engine. The standard augmentation suite applied during training encompasses rigorous spatial and intensity-level variations, explicitly comprising random flipping, joint rotation-scaling, additive gray-scale brightness shifts, contrast manipulation, gamma modifications, and randomized Gaussian noise injection. The scale augmentation range directly determines the viable dynamic spacing bounds ($\alpha$) exploited during inference. No test-time augmentation (TTA) was applied during the inference and deployment phases to ensure minimal online operational latency.

For production deployment, the trained PyTorch models were converted to the OpenVINO intermediate representation format (.bin/.xml), enabling CPU-optimized inference on the target Windows 10 Pro workstation (Intel Xeon Silver 4214R, 12 cores @2.40 GHz, 32 GB RAM).

### 3.2. Evaluation Metrics

To rigorously evaluate the proposed method, we utilize a complementary set of metrics assessing geometric overlap, clinical volume agreement, and computational efficiency.

**Geometric Accuracy:** The Dice Similarity Coefficient (DSC) is employed to quantify the spatial overlap between the predicted segmentations and ground-truth annotations.

**Clinical Volumetric Agreement:** In accordance with clinical measurement validation standards [11], Bland–Altman analysis is performed to assess the agreement between automated and manual volume measurements. For each anatomical structure, we report the mean bias and the 95% Limits of Agreement ($\text{LoA} = \text{bias} \pm 1.96 \cdot \text{SD}$). To demonstrate clinical utility, a stringent $\pm 10\%$ relative volumetric error bound is adopted as a reference threshold for acceptable clinical variance.

**Computational Efficiency:** To evaluate the feasibility of real-time clinical deployment, computational performance is benchmarked using the mean end-to-end inference wall-clock time (seconds) and peak memory consumption (GB) on a standard workstation.

### 3.3. Quantitative Segmentation Performance and Volumetric Agreement

We evaluated the performance of the finalized end-to-end pipeline on an independent deployment test cohort ($N = 105$). Notably, due to variations in standard clinical scan protocols, individual CT volumes did not universally encompass the full anatomic distribution of all ten target structures. Consequently, quantitative evaluation was strictly confined to volumes where a given structure was substantially covered, resulting in variable effective sample sizes ($N = 20{\sim}50$) per target, as detailed in Table 2. Specifically, macro-compartments such as subcutaneous fat and total muscles (Muscles_all) were strictly quantified within the standard abdominal region spanning from the dome of the liver to the iliac crest. The quantitative evaluation decoupling geometric accuracy and absolute/relative volumetric agreement are summarized below.

#### 3.3.1. Per-Structure Geometric and Volumetric Accuracy

Table 2 provides a comprehensive statistical breakdown of the segmentation accuracy and Bland–Altman agreement metrics across all target anatomical structures. The proposed model demonstrated exceptional geometric fidelity, with eight out of ten structures achieving a mean DSC $> 0.950$. The highest accuracy was observed in large, well-demarcated tissue compartments, namely subcutaneous fat (0.982 ± 0.011) and the quadriceps femoris (0.975 ± 0.017). Conversely, the lowest geometric performance occurred in the intercostal muscles (0.924 ± 0.025), which is inherently attributable to their thin, ribbon-like morphology and high surface-to-volume ratio that amplifies boundary-voxel misclassifications.

**Table 2. Quantitative segmentation accuracy and Bland–Altman volumetric agreement**

| **Target Structure** | N | **Mean DSC (± SD)** | **Relative Volumetric Bias [95% LoA] (%)** | **Absolute Volumetric Bias [95% LoA] (mL)** |
|---|---|---|---|---|
| Subcutaneous Fat | 50 | 0.982 ± 0.011 | 1.11 [-2.33, 4.56] | 13.2 [-34.0, 60.4] |
| Visceral Fat | 50 | 0.961 ± 0.020 | 0.59 [-7.42, 8.59] | -6.0 [-113.8, 101.8] |
| Muscles_all | 50 | | | |
| Erector Spinae | 28 | 0.973 ± 0.013 | 0.93 [-1.64, 3.50] | 8.1 [-15.7, 31.9] |
| Psoas Major | 20 | 0.969 ± 0.016 | 1.21 [-2.33, 4.76] | 3.6 [-5.9, 13.2] |
| Iliacus | 20 | 0.955 ± 0.018 | -0.31 [-6.58, 5.96] | -2.1 [-21.6, 17.4] |
| Quadriceps Femoris | 22 | 0.975 ± 0.017 | 0.27 [-1.25, 1.80] | 5.0 [-27.7, 37.8] |
| Biceps Femoris | 22 | 0.949 ± 0.037 | 2.22 [-7.26, 11.71] | 8.0 [-22.7, 38.7] |
| Pectoralis | 29 | 0.974 ± 0.015 | -1.47 [-4.13, 1.20] | -6.3 [-15.7, 3.1] |
| Intercostal | 29 | 0.924 ± 0.025 | 6.32 [-6.01, 18.65] | 23.8 [-18.0, 65.6] |

Bland–Altman analysis revealed tight limits of agreement and negligible systemic biases for both absolute and relative volume estimations across most structures. For the majority of major muscular and adipose compartments, the 95% LoA for relative volume errors fell comfortably within a narrow $\pm 10\%$ clinical acceptance window, validating the clinical reliability of the automated derivations. Specifically, structures such as the erector spinae, psoas major, iliacus, and quadriceps femoris all exhibited highly tightly bounded relative LoAs.However, marginal variances beyond the strict $\pm 10\%$ limit were noted in the remaining two structures: the biceps femoris (95% LoA: $[-7.26\%, 11.71\%]$) and the intercostal muscles (95% LoA: $[-6.01\%, 18.65\%]$). This expanded interval does not signify algorithmic instability, but rather represents a mathematical artifact common to thin-line structures; their exceptionally small absolute baseline volumes inherently amplify the mathematical penalty of minor, voxel-level boundary displacements during relative error calculations.

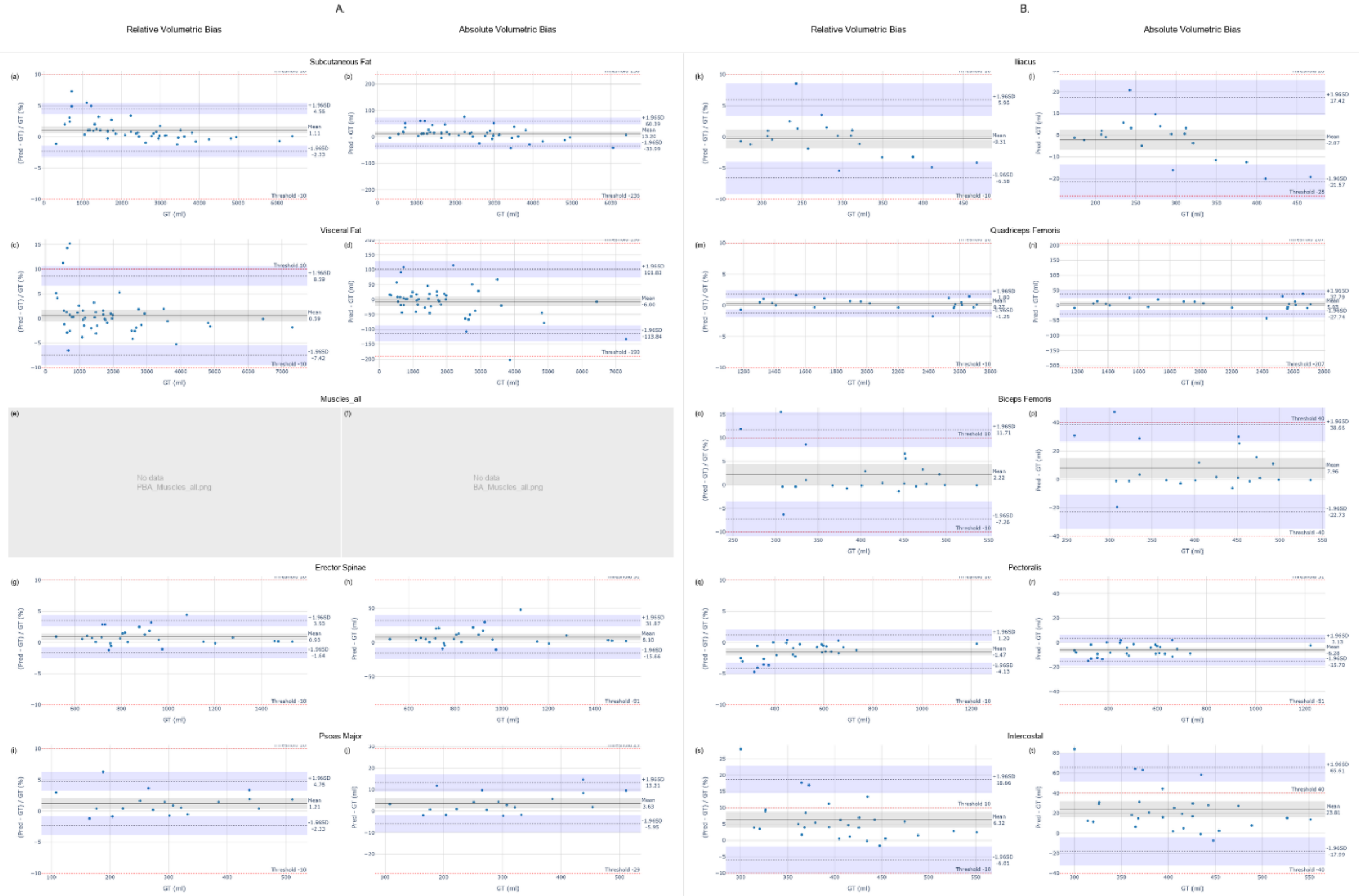

Figure 3. Bland-Altman analysis of automated segmentation performance for body composition compartments from standing CT. A. shows relative volumetric bias and absolute volumetric bias (BA, mL) for subcutaneous adipose tissue (SAT), visceral adipose tissue (VAT), Muscles_all, erector spinae, and psoas major. B. shows the same analyses for iliacus, quadriceps femoris, biceps femoris, pectoralis, and intercostal muscles. In each subplot, the solid black line indicates the mean bias and dashed black lines represent the 95% LoA. Subfigures are labeled (a)–(t).

### 3.3.2. Benchmark Comparison with TotalSegmentator

To demonstrate the superior efficacy of the proposed model, a comprehensive head-to-head comparison was conducted against the widely recognized benchmark TotalSegmentator on overlapping target structures.

As detailed in the comparative analysis (Table 3), the proposed model consistently outpaced TotalSegmentator across all evaluation metrics, yielding statistically significant improvements in both geometric fidelity (DSC) and volumetric agreement (all $p < 0.05$, with most $p < 0.001$). This performance gap is most pronounced in the erector spinae muscle, where TotalSegmentator exhibited a notable drop in geometric accuracy (mean DSC: 0.886 ± 0.013) and a massive underestimation of volume (relative bias: -15.89%; absolute bias: -146.6 mL; both $p < 0.001$). Conversely, the proposed model demonstrated exceptional precision in tracing visceral adipose tissue (VAT) along the highly intricate intestinal boundaries, yielding far more tightly conformed contours than the benchmark.

Visual inspection of the segmentation masks revealed that TotalSegmentator tends to generate overly smoothed, generalized predictions that fail to conform tightly to the complex, irregular anatomical boundaries of the muscular fascial planes. This severe "oversmoothing" artifact inherently causes voxel misclassifications at the tissue interfaces, accumulating into the severe geometric and volumetric discrepancies quantified above.

Furthermore, in the quantification of subcutaneous adipose tissue (SAT), our model successfully isolated true SAT by strictly excluding intermuscular adipose tissue (IMAT)

**Table 3. Comparative analysis of segmentation accuracy and Bland–Altman volumetric agreement between the proposed model and TotalSegmentator.**

| **Methods** | **Target Structure** | N | **Mean DSC (± SD)** | **Relative Volumetric Bias [95% LoA] (%)** | **Absolute Volumetric Bias [95% LoA] (mL)** |
|---|---|---|---|---|---|
| TotalSegmentator | Subcutaneous Fat | 50 | 0.946 ± 0.037 | 2.86 [-6.68, 12.41] | 85.3 [-165.2, 335.9] |
| | Visceral Fat | 50 | 0.929 ± 0.027 | 2.17 [-7.68, 12.02] | 40.1 [-134.4, 214.6] |
| | Erector Spinae | 28 | 0.886 ± 0.013 | -15.89 [-20.18, -11.61] | -146.6 [-250.8, -42.3] |
| Ours | Subcutaneous Fat | 50 | **0.982 ± 0.011 *** ** | **1.11 [-2.33, 4.56] * ** | **13.2 [-34.0, 60.4] *** ** |
| | Visceral Fat | 50 | **0.961 ± 0.020 *** ** | **0.59 [-7.42, 8.59] * ** | **-6.0 [-113.8, 101.8] *** ** |
| | Erector Spinae | 28 | **0.973 ± 0.013 *** ** | **0.93 [-1.64, 3.50] *** ** | **8.1 [-15.7, 31.9] *** ** |

*Note: Bold values indicate better performance. * $p < 0.05$, ** $p < 0.01$, *** $p < 0.001$ compared with TotalSegmentator using Wilcoxon signed-rank test.*

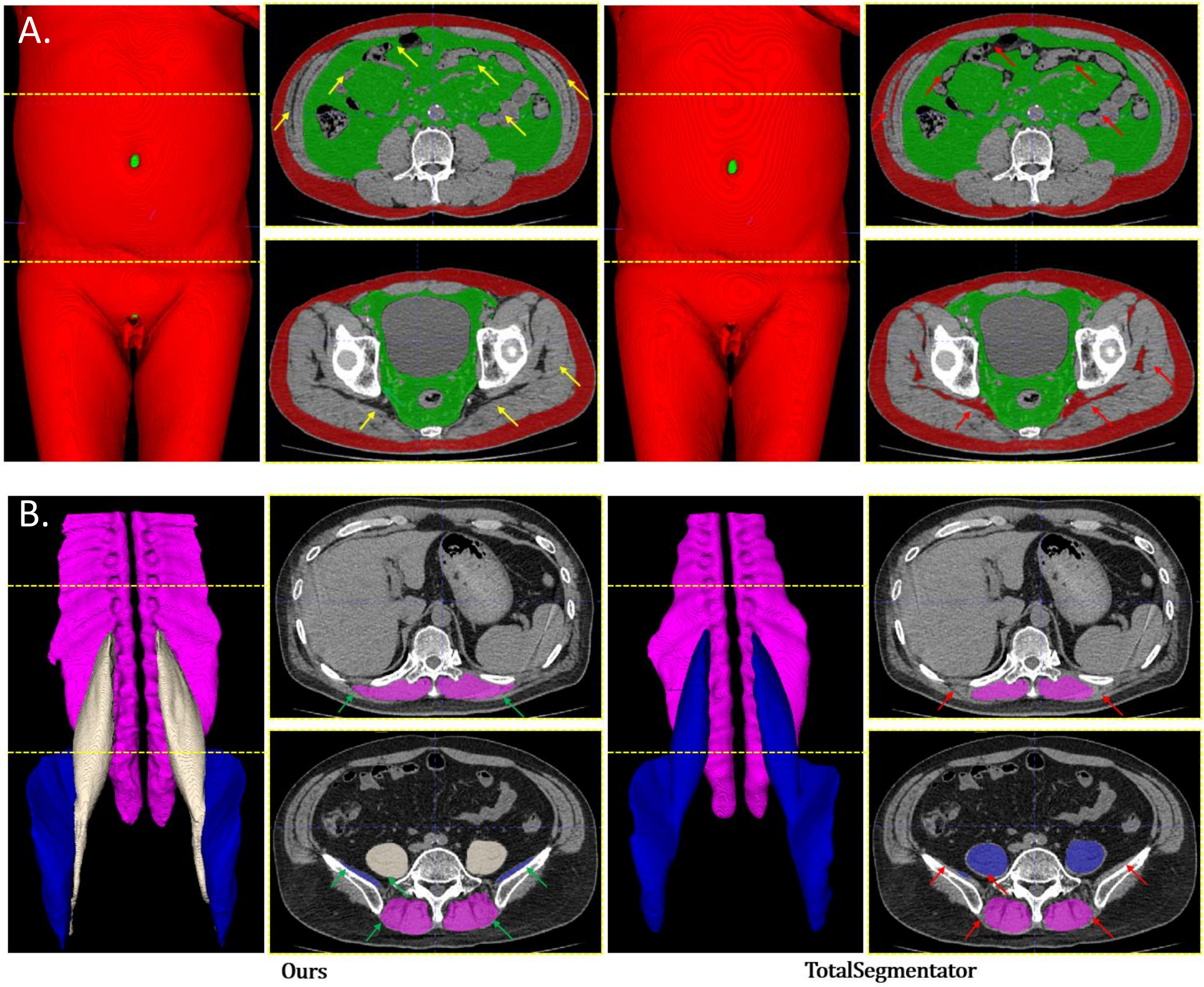

Figure 4. Visual comparison of segmentation boundaries between TotalSegmentator and the proposed model. A. SAT and VAT. the proposed model achieves tighter boundary conformity for visceral fat and accurately delineates subcutaneous fat by excluding intermuscular adipose tissue. B. Erector Spinae and Iliapsoas. TotalSegmentator exhibits a pronounced "oversmoothing" artifact and conservative segmentation, failing to conform to the anatomical boundaries of the muscular fascial planes (arrows). In contrast,

### 3.4. Computational Efficiency and Ablation Analysis

#### 3.4.1. Inference Performance Profile

To evaluate clinical deployment feasibility on standard hospital hardware, the end-to-end pipeline was benchmarked on a CPU-only workstation (Intel Xeon Silver 4214R CPU, 12-core, OpenVINO runtime, 32 GB RAM). The finalized frozen pipeline achieved an average end-to-end inference wall-clock time of 44.5 seconds per volume, with a peak memory of 4.73 GB, safely rendering it compatible with routine clinical workflows.

#### 3.4.2. Ablation of Acceleration Modules

To dissect the isolated empirical contributions of each architectural and post-processing optimization module toward inference speedup and memory suppression, a controlled ablation study was conducted.

**Table 4. Ablation analysis of acceleration and memory optimization modules on the test cohort.**

| Configuration | Mean DSC | Inference Time (s) | Peak Memory (GB) |
|---|---|---|---|
| Baseline *(Fixed Spacing and patch + Sliding Window)* | **0.964** | 74.0 | 6.40 |
| + Topology-Aware Asymmetric Resampling | **0.964** | 62.2 | 5.11 |
| + Anisotropic Patching with Dynamic Spacing | 0.962 | 45.1 | 5.20 |
| + Group Inference | 0.962 | **44.5** | **4.73** |

As detailed in Table 4, the sequential integration of the proposed acceleration and memory optimization components progressively minimized computational overhead and memory footprints while maintaining a stable, high-fidelity segmentation accuracy (Mean DSC remained decoupled from efficiency trade-offs, fluctuating marginally between 0.962 and 0.964)

The native Baseline configuration (utilizing standard isotropic spacing and vanilla sliding-window inference) inherently suffers from heavy computational and volatile memory loads due to the dense processing of redundant backgrounds and sub-volume overlaps.

Implementing Topology-Aware Asymmetric Resampling achieved the first major acceleration (62.2s), as it confines intensive interpolations strictly to tissue boundaries rather than the uniform interiors. The runtime was further compressed to 45.1s by introducing Anisotropic Patching with Dynamic Spacing, which adapts to the input voxel anisotropy and drastically reduces the total number of patch inputs. Finally, integrating the Group Inference mechanism successfully resolved the peak memory bottleneck, driving it down to its optimal nadir of 4.74GB (with a fast runtime of 44.5) via an incremental local window processing and on-the-fly sub-volume release protocol. These results firmly validate the efficiency and feasibility of deploying our heavy multi-structure framework on standard GPU-free CPU workstations.

## 4. Discussion and Conclusion

### 4.1. Methodological and Practical Implications

The empirical results demonstrate that deep learning-based body composition analysis can be integrated into clinical workflows without requiring dedicated GPU infrastructure. While model performance is fundamentally linked to network capacity, optimizing inference execution parameters allows high-throughput multi-structure segmentation within standard computational constraints.

A key characteristic of the proposed framework is that its architectural optimizations are preserved during deployment without altering the underlying network representations. The Dynamic Spacing and Anisotropic Patching strategy capitalizes on the network's spatial scale tolerance—established via spatial data augmentations during training—to dynamically adapt the sliding-window aspect ratio during inference, thereby minimizing redundant patch overlap. Similarly, the Group Inference protocol and Topology-Aware Asymmetric Resampling adjust the sequence of tensor operations and resolution recovery without modifying the prediction logits. Consequently, these computational reductions avoid the loss of geometric accuracy often associated with lossy compression techniques, such as model quantization or knowledge distillation.

The Group Inference mechanism addresses the memory overhead caused by accumulating high-dimensional softmax tensors during multi-class 3D segmentation. In extensive scan ranges covering ten target structures, the memory required for accumulated logits scales with volume size and class count, often exceeding the network's activation memory. By continuously tracking completed spatial regions and converting finalized tensors into compact integer label maps on-the-fly, the pipeline limits peak memory accumulation to 4.73 GB. This reduction permits deployment on standard hospital workstations lacking discrete graphics hardware.

4.2. Limitations and Failure Cases

Several limitations remain to be addressed in future iterations:

Volumetric Sensitivity in Intercostal Muscles: The intercostal muscles exhibited the lowest segmentation performance ($0.924 \pm 0.025$) and a wider relative 95% LoA ($[-6.01\%, 18.65\%]$), which did not fully satisfy the prospective $\pm 10\%$ relative volumetric error criterion. This variance is primarily a consequence of the structure's thin, parallel morphology and small absolute baseline volume. Under these geometric conditions, sub-voxel boundary misclassifications represent a disproportionately large percentage of the total structure volume, mathematically amplifying the relative error. Therefore, automated measurements of the intercostal muscles should be interpreted with caution in clinical tasks requiring high absolute precision, such as specific respiratory muscle assessments.

Hardware Evaluation Constraints: The evaluation focuses on CPU-only execution (OpenVINO runtime on Intel Xeon) to match typical clinical deployment environments, the processing advantages and scalability of this pipeline in GPU-accelerated multi-user configurations have not been evaluated.

Cohort Size and Longitudinal Validation

Finally, although the independent test set ($N = 105$) incorporates diverse scanner vendors and acquisition parameters, a larger multi-center cohort is required to establish broader statistical boundaries. Longitudinal evaluation was also restricted by the available follow-up sample size. Expanding the serial monitoring datasets is necessary to further verify the system's sensitivity to minor, long-term volumetric variations associated with progressive sarcopenia.

4.3. Conclusion

This work presents a coarse-to-fine hierarchical framework optimized for the multi-structure segmentation of muscle and adipose tissue compartments from CT data. By combining a dynamic spacing and anisotropic patching strategy with an incremental group inference protocol and topology-aware resampling, the pipeline reduces peak memory consumption and post-processing latency during CPU execution without altering prediction logits or compromising geometric accuracy. Experimental evaluation on an independent test cohort confirms that the system maintains high spatial overlap metrics and achieves acceptable volumetric agreement for eight out of ten target structures under real-world data heterogeneity. While small-volume, high-surface-area structures like the intercostal muscles remain sensitive to boundary misclassifications, the framework altogether provides a computationally viable and mathematically stable paradigm for automated, large-scale body composition analysis on standard clinical workstations.